\documentclass[aps,prd,superscriptaddress,onecolumn,showpacs,showkeys]{revtex4}

\usepackage{amssymb}
\usepackage{amsmath}
\usepackage{epsfig}
\usepackage{eurosym}
\usepackage{amsfonts}
\usepackage{array}
\usepackage{amsthm}
\usepackage{bm}
\usepackage{palatino}
\usepackage{changes}
\usepackage{supertabular}
\usepackage{graphics}
\usepackage{color}
\usepackage{graphicx}
\usepackage[utf8]{in putenc}
\usepackage{cancel}
\usepackage{ulem}
\usepackage{enumerate}
\usepackage{mathtools}
\usepackage{pgfplots}
\usepackage{subfig}
\usepackage{float}
\usepackage{here}
\usepackage{mathrsfs}
\usepackage{dcolumn}
\usepackage{tikz}\usetikzlibrary{calc}
\usetikzlibrary{decorations.pathreplacing}
\usepackage[colorlinks=true,linkcolor=blue,citecolor=red]{hyperref}
\newcommand{\be}{\begin{equation}}
\newcommand{\ee}{\end{equation}}
\newcommand{\bea}{\setlength\arraycolsep{2pt} \begin{eqnarray}}
\newcommand{\eea}{\end{eqnarray}}

\def\0{{\sst{(0)}}}
\def\1{{\sst{(1)}}}
\def\2{{\sst{(2)}}}
\def\3{{\sst{(3)}}}
\def\4{{\sst{(4)}}}
\def\5{{\sst{(5)}}}
\def\6{{\sst{(6)}}}
\def\7{{\sst{(7)}}}
\def\8{{\sst{(8)}}}
\def\sst#1{{\scriptscriptstyle #1}}

\begin{document}

\title{ Weak Deflection angle of some classes of non-linear electrodynamics black holes via Gauss-Bonnet Theorem}

\author{H.  El Moumni} 
\email{hasan.elmoumni@edu.uca.ma}
\affiliation{EPTHE, Physics Department, Faculty of Science,  Ibn Zohr University, Agadir, Morocco.}

\author{K.  Masmar} 
\email{karima.masmar@edu.uca.ac.ma}
\affiliation{ Laboratory of  High Energy Physics and Condensed Matter
Hassan II University,  Faculty of Science Ain Chock, Casablanca, Morocco. }

\author{Ali \"{O}vg\"{u}n}
\email{ali.ovgun@emu.edu.tr}
\homepage[]{https://www.aovgun.com}
\affiliation{Physics Department, Faculty of Arts and Sciences, Eastern Mediterranean
University, Famagusta, North Cyprus, via Mersin 10, Turkey}

\begin{abstract}

In this paper, we study the gravitational lensing by some black hole classes within the non-linear electrodynamics in weak field limits. First, we calculate an optical geometry of the non-linear electrodynamics black hole  then we use the Gauss-Bonnet theorem for finding deflection angle in weak field limits.   The effect of non-linear electrodynamics on the deflection angle in leading order terms is studied. Furthermore, we discuss the effects of the plasma medium on the weak deflection angle.

\end{abstract}
\keywords{Weak Deflection Angle; Gauss-Bonnet theorem; Nonlinear electrodynamics; Gravitational lensing.}
\pacs{ 04.70.Dy, 95.30.Sf, 97.60.Lf }

\date{\today}

\maketitle

\section{Introduction}

Einstein's general theory of relativity \cite{Einstein:1916vd} is considered as the most beautiful and elegant theory of the last century. This theory has passed several tests with spectacular successes showing this compatibility with experimental observations since 1920 \cite{Dyson:1920cwa}, when   Dyson, Eddington and Davidson have measured the gravitational bending of light by the Sun. A quote of years ago, other huge experiments also unveil the predictions of such a theory one can cite the detection of the gravitational waves \cite{TheLIGOScientific:2016agk}  and the realization of the first black hole image in M87 galaxy announced by Event Horizon Telescope (EHT) \cite{Abbott:2016blz,Akiyama:2019cqa}. One of the key phenomena predicted by the gravitation theory is the "gravitational lensing" which can be interpreted by the ability of gravity to bend light. Such phenomena have been observed many times and it has been established as one of the powerful tools in astronomy and cosmology especially for investigation of the dark sector of our universe \cite{Bartelmann:1999yn}- \cite{Er:2013efa}.

As of late, Gibbons and Werner \cite{Gibbons:2008rj} changed the standard perspective identified with the manner in which we ordinarily
evaluate the deflection angle. They demonstrated that one can compute the deflection angle in a very elegant manner; Specifically, they have utilized the Gauss-Bonnet theorem (GBT) with regards to optical geometry. It is important to note that in this method, the deflection of photon can be seen as a global effect. One can focus only on a nonsingular area outside of the photon ray. This method is applied on various black holes \cite{Crisnejo:2019ril}-\cite{Jusufi:2018jof}  and wormholes \cite{Ovgun:2018xys,Javed:2019qyg,Jusufi:2017mav,Jusufi:2018kmk,Ovgun:2018prw,OvgunUniverse,Ovgun:2018fnk,Li:2019vhp,Goulart:2017iko,Ono:2018ybw} to find the weak deflection angle.

In the framework of Gravitation theory, one can discover peculiarity of singularity–free solutions of the Einstein field equations coupled to suitable nonlinear electrodynamics (NLED), which in the weak field limit reduces to the ordinary linear Maxwell theory.
In in this context NLED charged black hole solutions were derived and discussed in a  number  of  works  \cite{Allahyari:2019jqz}-\cite{Nam:2018tpf}.
Initially, the NLED was elaborated by Born and  Infeld’s to expel the central singularity of  a  point charge  and the related energy divergence by generalizing Maxwell’s theory  \cite{Born:1934gh}. Then, the enthusiasm for NLED in current studies is  to an enormous degree roused by 
 the discovery  that  some  kinds  of  NLED  appear  as  limiting  cases  of  certain  models  of string theory \cite{Tseytlin:1999dj,Seiberg:1999vs}.

Here, the main aim of the paper is to contribute to such interest area of gravitational physics by recalling the Gauss-Bonnet theorem to evaluate the the deflection angle of two class of black hole solutions of gravity coupled to nonlinear electrodynamics. These calculations are elaborated in the vacuum and by considering a plasma medium.

This paper is organized as follows, after a concise introduction of the black hole solutions we compute their optical metrics and the Gaussian optical curvatures. After what and by the help of the GBT, the deflection angle of light for such black holes configuration is computed. In section Sec.3, we observe the graphically the variation bahaviour of the deflection angle in non-plasma and plasma medium and make a comparison between both mediums, then we briefly summarize our results.

\section{ Computation of weak lensing by non linear electrodynamics black hole within the Gauss-Bonnet theorem}

Our starting point is the following action in which gravity is coupled to nonlinear
electrodynamics fields \cite{Sajadi:2019hzo,Nam:2018tpf}
\begin{equation}
S=\frac{1}{16\pi }\int d^4x \sqrt{-g}\left[ R- 4\pi\mathcal{L}(F)\right],
\end{equation}
where $R$ denotes the scalar curvature and $\mathcal{L}(F)$ is a  function of the invariant  $F:= \frac{1}{4}F_{\mu\nu}F^{\mu\nu}$  and $\tilde{F}:= \frac{1}{4}F_{\mu\nu}\star F^{\mu\nu}$ built form Faraday tensor ${\bf F}= \frac{1}{2}F_{\mu\nu}d x^\mu \wedge dx^\nu$ and its Hodge dual $\star {\bf F}$. In this paper,  we will deal with the non-linear electrodynamics terms $\mathcal{L}(F)$ are explicitly given by  \cite{Sajadi:2019hzo,Nam:2018tpf}

\begin{equation}
\mathcal{L}(F)=\left\{\begin{array}{cc} \frac{8 M^3 F (6M(-2F)^{\frac{1}{4}} +\sqrt{Q}- 2 M Q (-2F)^{\frac{3}{4}}}{\sqrt{Q}(2M +Q^{\frac{3}{2}} (-2F)^{\frac{1}{4}}+2 M Q (-2F)^{\frac{1}{2}}  )^3 }
 & \tt{NLED 1} \\\frac{3M}{|Q|^3}\frac{(2Q^2 F)^{3/2}}{[1+(2Q^2 F)^{3/4}]^2}
& \tt{NLED 2},\end{array}\right.
\end{equation}
in which $M$ and $Q$ are the mass and charge of black holes respectively. Herein, we would like to deal with  static and spherical symmetric  black hole given by  the following ansatz
\begin{equation}\label{lineelement}
dt^2=-f(r)dt^2+\frac{dr^2}{f(r)}+r^2(d\theta^2+\sin \theta d \phi^2 ),
\end{equation}
where the metric functions written as \cite{Sajadi:2019hzo,Nam:2018tpf}
\begin{equation}
f(r)=\left\{\begin{array}{cc}1-\frac{2 M r}{r^2+\frac{Q^2}{2 M}r+Q^2} & \tt{NLED 1} \\1-\frac{2 M r^2}{r^3+Q^3}  & \tt{NLED 2}.\end{array}\right.
\end{equation}

 In order to consider the Gauss-Bonnet theorem \cite{Gibbons:2008rj}, we introduce the 
 optical space simply written in equatorial plane $(\theta=\frac{\pi}{2})$ to get null geodesics ($ds^2 = 0$) as 
\begin{equation}
dt^2=g_{ij}^{opt}dx^idx^j=\frac{dr^2}{f(r)^2}+\frac{r^2d\phi}{f(r)}.
\end{equation}
The Gaussian optical curvature $\mathcal{K}$ which is an intrinsic property of spacetime and which  corresponds to the optical metric is related to its Ricci scalar $\mathcal{R}$ via the following formula
\begin{equation}
\mathcal{K}=\frac{\mathcal{R}}{2},\quad \text{where}\quad \mathcal{R}=\frac{-f'(r)^2}{2}+f''(r)f(r).
\end{equation}

Recalling the both expressions of $f(r)$, the Gaussian optical curvature is given explicitly by 

\begin{equation}\label{Gaussian}
\mathcal{K}\approx\left\{\begin{array}{cc}3\,{\frac {{Q}^{2}}{{r}^{4}}}+ \left( -2\,{r}^{-3}+6\,{\frac {{Q}^{2}
}{{r}^{5}}} \right) M
& \tt{NLED 1} \\ \left( -2\,{r}^{-3}+20\,{\frac {{Q}^{3}}{{r}^{6}}} \right) M & \tt{NLED 2}.\end{array}\right.
\end{equation}

Now, by utilizing Gauss-Bonnet Theorem (GBT), we evaluate the deflection angle of photon by such black holes.
 Using GBT in the region $\mathcal{G}_R$, given as
\begin{equation}
\int_{\mathcal{G}_R}\mathcal{K}dS+\oint_{\partial\mathcal{G}_R}k dt+\sum_t\hat{\alpha}=2\pi \mathcal{X}({\mathcal{G}_R})
\end{equation}
in which, $k$ represent the geodesic curvature, $\mathcal{K}$ is for the Gaussian optical curvature as well as the geodesic curvature $k$ is written as $k=\bar{g}(\nabla_{\hat{\alpha}}\hat{\alpha},\check{\alpha})$ so that $\bar{g}(\hat{\alpha},\hat{\alpha})=1$, in which $\hat{\alpha}$ is the unit acceleration vector and $\alpha_t$ is for the exterior angle at $t^{th}$ vertex respectively. For $R\rightarrow \infty$,the jump angles equal to $\pi/2$, so that $\alpha_O+\alpha_S\rightarrow \pi$. Here, $\mathcal{X} (\mathcal{G}_R ) = 1$ stands for a Euler characteristic number and $\mathcal{G}_R$ is the nonsingular domain. Then we find that
\begin{equation}
\int\int_{\mathcal{G}_R}\mathcal{K}dS+\oint_{\partial\mathcal{G}_R}k dt+\hat{\alpha}=2\pi \mathcal{X}({\mathcal{G}_R})
\end{equation}
where, the total jump angle is $\hat{\alpha}=\pi$, When $R\rightarrow \infty$, then the remaining part is $k{D_R}=|\nabla_{\dot{D}_R}\dot{D}_R|$
radial component for geodesic curvature:
\begin{equation}\label{eq11}
(\nabla_{\dot{D}_R}\dot{D}_R)^r=\dot{D}^\phi_R\partial_\phi\dot{D}^r_R+\Gamma^r_{\phi\phi}(\dot{D}^\phi_R)^2
\end{equation}
At very high $R$, $D_R:=r(\phi)=R=const$. Hence,  we write that the leading term of equation Eq.\eqref{eq11} vanishes and $(\dot{D}^\phi_R)^2=\frac{1}{f(r^\star)}$. Recalling $\Gamma^r_{\phi\phi}=\frac{2f(r)-rf'(r)}{2rf(r)}$, we get $(\nabla_{\dot{D}_R}\dot{D}_R)^r=\frac{1}{R}$ and which proves that the geodesic curvature is not effected to the topological defects. We can write $dt=Rd\phi$. Thus; $k(D_R)dt=d\phi$.
From the previous results, we get
\begin{equation}
\int\int_{\mathcal{G}_R}\mathcal{K}dS+\oint_{\partial\mathcal{G}_R}k dt
\overset{R\rightarrow\infty}{=} \int\int_{S_\infty}\mathcal{K}dS+\int_0^{\pi+\alpha}d\phi.
\end{equation}
At $0^{th}$ order, weak field deflection limit of the light is defined as $r(t)=b/\sin\phi$. Consequently , the deflection angle can now expressed as
as:
\begin{equation}\label{15}
\alpha=-\int_0^\pi\int_{b/\sin\phi}^\infty \mathcal{K}\sqrt{\det\bar{g}} du d\phi.
\end{equation}

here we put the leading term of Eq.\eqref{Gaussian}  into above Eq.\eqref{15}, so the obtained deflection angle up to leading order term is computed as:
\begin{equation}
\alpha\approx\left\{\begin{array}{cc} 4
\,{\frac {M}{b}}-3/4\,{\frac {{Q}^{2}\pi}{{b}^{2}}}-8/3\,{\frac {{Q}^{2}M}{{b}^{3}}}
& \tt{ NLED1} \\4\,{\frac {M}{b}}  -{\frac {15\,M\pi\,{Q}^{3}}{8\,{b}^{4}}}     & \tt{NLED2}\end{array}\right.
\end{equation}
 While  we can  immediately see that the charge term decreases the deflection angle in the both classes.

\bigskip

Lensing in a vacuum does not involve dispersive proprieties of a photon. Furthermore, lenses are usually besieged \cite{Er:2013efa} by plasma, which discloses a nontrivial component for the deflection angle. Gravitational deflection in a plasma medium encourages refraction including more deflection such that the information is encoded in the refraction index \cite{Bisnovatyi-Kogan:2017kii}.

In order to encompass the effects of plasma, we suppose that the photon travels from vacuum to a hot, ionized gas medium, where the $v$ is the speed of light through the plasma. Then we can write the refractive index, $n(r)$ as follows:
\begin{equation}
n(r)\equiv\frac{v}{c}=\frac{1}{dr/dt},\quad\quad \{\because c=1\}.
\end{equation}
The refractive index $n(r)$ for such black hole is obtained as \cite{Crisnejo:2019ril}
\begin{equation}
n(r)=\sqrt{1-\frac{\omega_e^2}{\omega_\infty^2}f(r)},
\end{equation}
where $\omega_e$ is the electron plasma frequency, and $\omega_\infty$ is the photon frequency measured by an observer at infinity. The line element figured in equation Eq.\eqref{lineelement} take a new form given  by
\begin{equation}
dt^2=g_{ij}^{opt}dx^idx^j=n^2(r)\left[\frac{dr^2}{f(r)^2}+\frac{r^2d\phi}{f(r)}\right].
\end{equation}


Taking account into the plasma medium the Gaussian optical curvatures Eq.\eqref{Gaussian} become
\begin{equation}\label{Gaussianplasma}\small
\mathcal{K}_{p}\approx\left\{\begin{array}{c}  5\,{\frac {{Q}^{2}{\omega_e}^{2}}{{r}^{4}{\omega_\infty}^{2
}}}+3\,{\frac {{Q}^{2}}{{r}^{4}}}-5\,{\frac {M{Q}^{2}{\omega_e
}^{2}}{{\omega_\infty}^{2}{r}^{5}}}+6\,{\frac {{Q}^{2}M}{{r}^{5}}}-
3\,{\frac {M{\omega_e}^{2}}{{\omega_\infty}^{2}{r}^{3}}}-2
\,{\frac {M}{{r}^{3}}} \quad\tt{NLED 1} \\ 36\,{\frac {M{Q}^{3}{\omega_e}^{2}}{{r}^{6}{\omega_\infty}^
{2}}}+20\,{\frac {M{Q}^{3}}{{r}^{6}}}-3\,{\frac {M{\omega_e}^{
2}}{{r}^{3}{\omega_\infty}^{2}}}-2\,{\frac {M}{{r}^{3}}}-156\,{
\frac {{M}^{2}{Q}^{3}{\omega_e}^{2}}{{r}^{7}{\omega_\infty}
^{2}}}-36\,{\frac {{M}^{2}{Q}^{3}}{{r}^{7}}}+12\,{\frac {{M}^{2}{
\omega_e}^{2}}{{r}^{4}{\omega_\infty}^{2}}}+3\,{\frac {{M}^
{2}}{{r}^{4}}}\quad \tt{NLED 2}\end{array}\right.
\end{equation}

By the help of the Gauss-Bonnet-theorem, one can express the deflection angle of such black hole classes within the plasma medium as

\begin{equation}\label{alphaplasma}
\alpha_{plasma}\approx\left\{\begin{array}{cc} 4\,{\frac {M}{b}}-8/3\,{\frac 
{{Q}^{2}M}{{b}^{3}}}-3/4\,{\frac {{Q}^{2}\pi}{{b}^{2}}}
-5/4\,{\frac {{Q}^{2}{\omega_e}^{2}\pi}{{b}^{2}{\omega_{\infty}}^{2}}}+{\frac {20\,M{Q}^{2}{
\omega_e}^{2}}{9\,{b}^{3}{\omega_{\infty}}^{2}}}+6\,{\frac {M{\omega_e}^{2}}{b{\omega_{\infty}}^{2}}}
& \tt{ NLED1} \\4\,{\frac {M}{b}}-{\frac {15\,M{Q}^{3}\pi}{8\,{b}^{4}}}+
 {\frac {27\,M{Q}^{3}{\omega_e}^{2}\pi}{8\,{b}^{4}{\omega_{\infty}}^{2}}}+6\,{\frac {M{\omega_e}^{2}}{b{\omega_{\infty}}^{2}}}
 & \tt{NLED2}
   \end{array}\right.
\end{equation}

\section{Conclusion}

\begin{figure}[!ht]
		\begin{center}
		\centering
			\begin{tabbing}
			\centering
			\hspace{9.3cm}\=\kill
			\includegraphics[scale=.8]{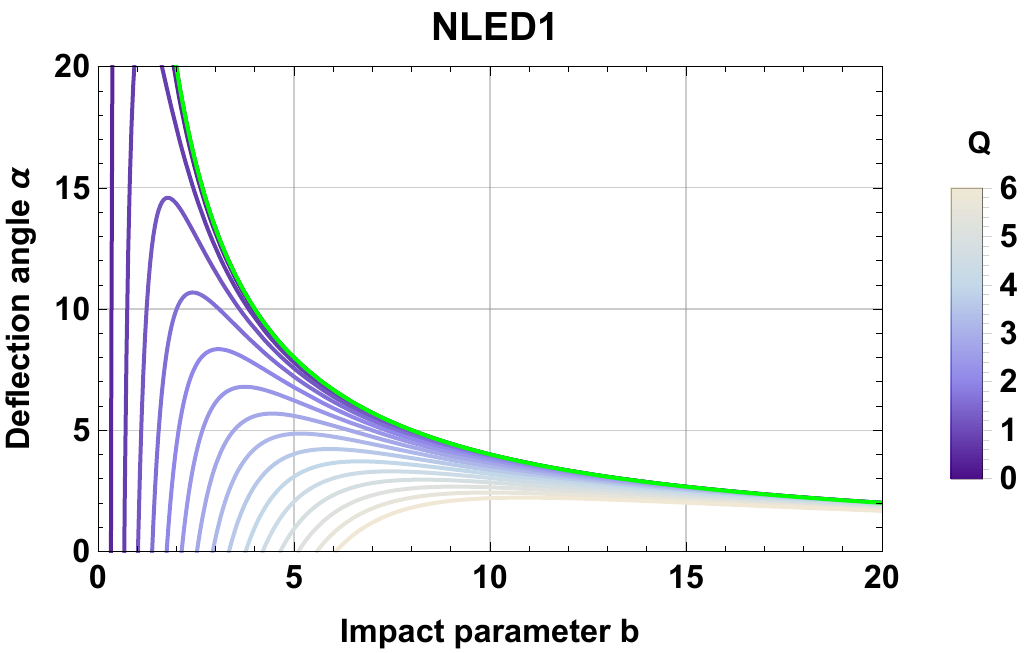} \>
			\includegraphics[scale=.8]{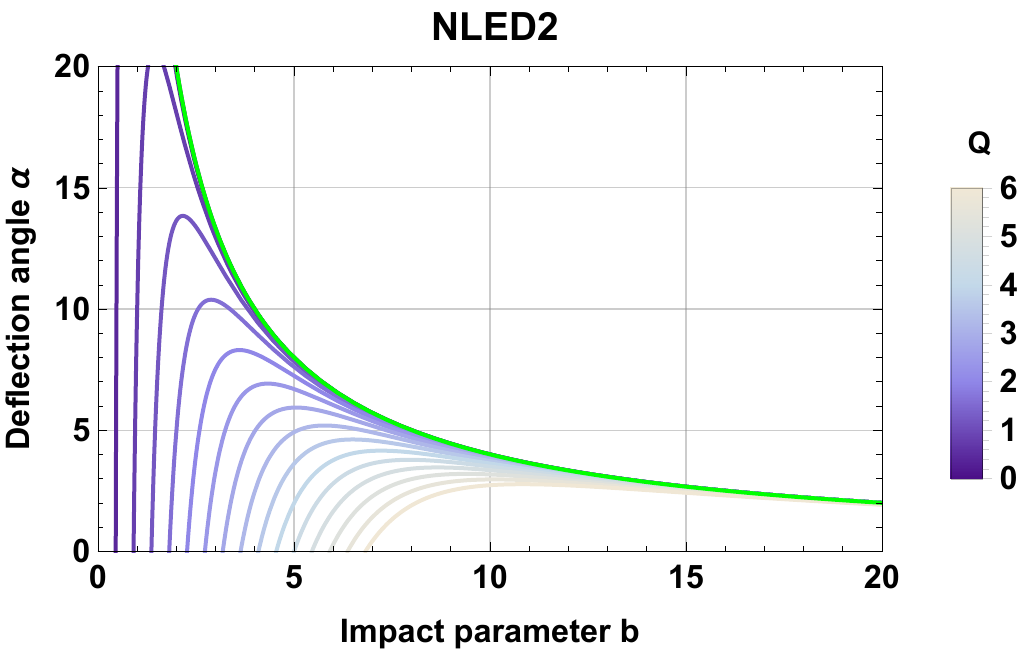} \\
		   \end{tabbing}

\caption{{\it \footnotesize  The relation between the deflection angle $\alpha$ and the parameter impact $b$ for each black hole class, within different values of the charge $Q$. The green line is associated with the vanishing charge $Q=0$. In both panels, we have set $M=10$.  }}\label{defvac}
\end{center}
\end{figure}

Having obtained the expressions of the deflection angle for each black hole classes in the vacuum and  plasma medium and to illustrate graphically  the the effect on nonlinear electrodynamics on such quantity, we depict in the figure Fig.\ref{defvac} the variation of the deflection angle $\alpha$ versus  the impact parameter $b$

From Fig.\ref{defvac}, one can easily see that, for both varieties  of the nonlinear electrodynamics black hole solution, the deflection angle $\alpha$ presents a maximum at 
    \begin{equation}
    b_{max,1}=\frac{3 \pi  Q^2+\sqrt{512 M^2 Q^2+9 \pi ^2 Q^4}}{16 M} \qquad \text{and}\qquad  b_{max,2}=\frac{1}{2} \sqrt[3]{15 \pi } Q
    \end{equation}
     respectively,   then decrease gradually and go to positive infinity. Comparing the deflection angles $\alpha$, one can notice that the $\alpha$ associated with $NLED1$ class is more important than the $NLED2$ one.  Another important remark is that the maximum $b_{max,2}$ depends only on the charge while $b_{max,1}$ is controlled by the mass $M$ and the charge $Q$. In the vanishing limit of charge $Q=0$ (green line) we recover the Schwarzschild  \cite{} behaviour, where the deflection angle is initially exponentially decreasing and then goes to positive infinity.

\begin{figure}[!ht]
		\begin{center}
		\centering
			\begin{tabbing}
			\centering
			\hspace{9.3cm}\=\kill
			\includegraphics[scale=.8]{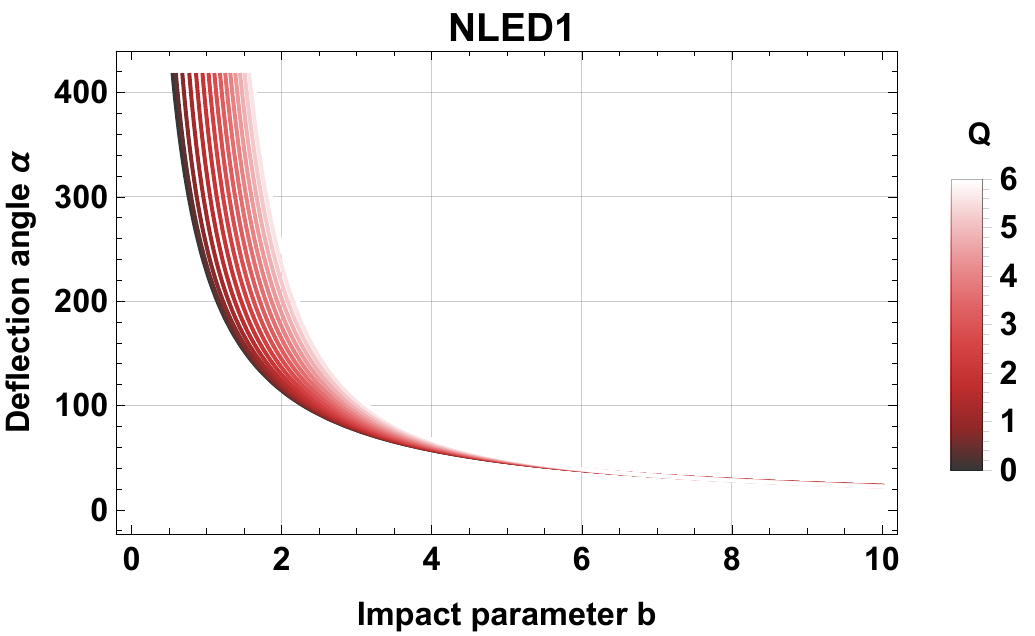} \>
			\includegraphics[scale=.8]{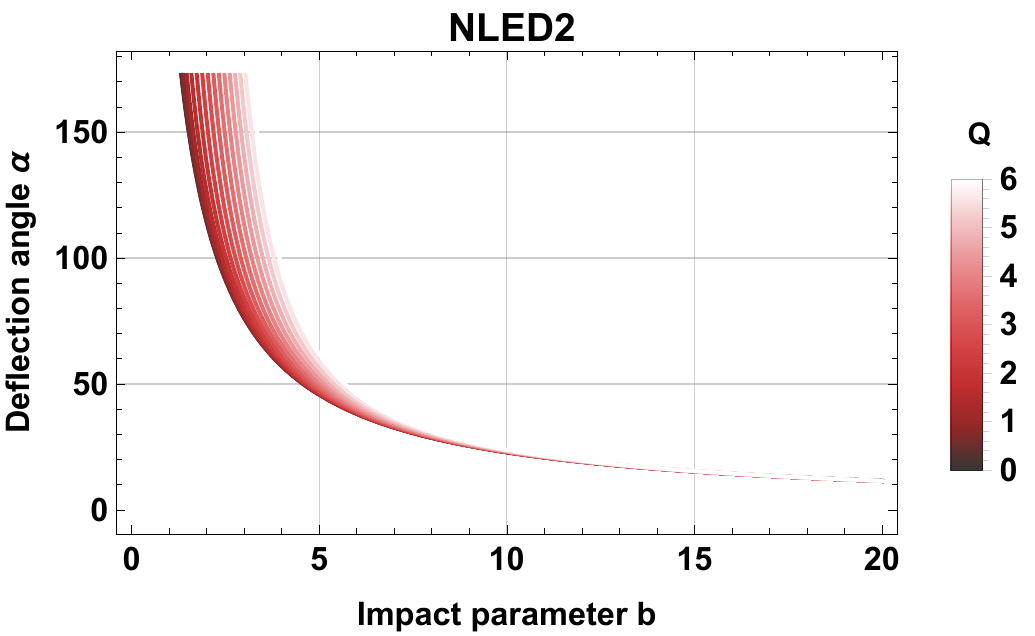} \\
		   \end{tabbing}
\caption{{\it \footnotesize  The relation between the deflection angle $\alpha$ and the parameter impact $b$ for each black hole class in the plasma medium, within  different values of charge $Q$. In both panels, we have set $M=10$.    }}\label{defmed}
\end{center}
\end{figure}

To unveil the effect of the plasma medium on the deflection angles of such black hole solutions, we depict in the figure Fig.\ref{defmed} the variation of $\alpha$ in terms of the parameter impact $b$. 

One can notice from this figure that the situation is quite different from the vacuum medium. Herein in the plasma medium, the maximum point disappears and the decreasing behaviour takes place: the deflection angle is initially exponentially decreasing and then goes to positive infinity. However, by the help of deflection angle expression Eq.\eqref{alphaplasma}, one can see that the value deflection angle increases within the introduction of the plasma medium.

In this work, we have investigated the weak gravitational lensing in the framework of nonlinear electrodynamics black hole solutions. For this end, we have considered the photon rays into the equatorial plane. Then we have calculated the associating optical metric and the Gaussian optical curvature.
By these quantities and the Gauss-Bonnet theorem we derive the expression of the deflection angle for each black hole background in vacuum and in plasma medium. We have also probed graphically the variation of the black hole deflection angle in terms of the impact parameter $b$ and within the charge and the plasma medium. We plan in the near future to extend our study using the Gauss-Bonnet theorem to other black hole  and dark object configurations.

\end{document}